\documentclass[%
 reprint,
 amsmath,amssymb,
 aps,
]{revtex4-2}

\usepackage{graphicx}
\usepackage{dcolumn}
\usepackage{bm}

\begin{document}

\preprint{APS/123-QED}

\title{Delicate Superconductivity in NaAlGe Single Crystal}

\author{Zhaoxu Chen$^{1,2}$}
\author{Yuxin Yang$^{1,3}$}
\author{Jun Deng$^{1}$}
\author{Tianping Ying$^{1}$}
\email{ying@iphy.ac.cn}
\author{Jianggang Guo$^{1,4}$}
\email{jgguo@iphy.ac.cn}
\author{Xiaolong Chen$^{1,4}$}
\email{chenx29@iphy.ac.cn}
\affiliation{ 
$^1$ \textit{Beijing National Laboratory  for Condensed Matter Physics, Institute of Physics, Chinese Academy of Sciences, Beijing 100190, China}}
\affiliation{
$^2$ \textit{School of Physical Sciences, University of Chinese Academy of Sciences, Beijing 100049, China}}
\affiliation{
$^3$ \textit{College of Materials Sciences and Opto-Electronic Technology, University of Chinese Academy of Sciences, Beijing 100049, China}}
\affiliation{
$^4$ \textit{Songshan Lake Materials Laboratory, Dongguan 523808, China}
}

\begin{abstract}
Nodal-line superconductor NaAlSi with a transition temperature ($T_c$) of 7 K has attracted considerable attention in recent years, whereas its Ge counterpart, NaAlGe, does not superconduct down to the lowest temperature regardless of their similar atomic and electrical structures. To tackle this enigma, we resort to the growth of NaAlGe single crystal and characterize its semimetallic ground state. Interestingly, when hole doped by oxidation, single-crystalline NaAlGe transforms from a semimetal to a superconductor ($T_c = 2.3  \sim$  3.5 K) with zero resistivity and a diamagnetic shielding fraction over 100$\%$, but without a thermodynamic response in heat capacity. Continuous x-ray diffractions reveal a transient new structure with a larger $c$ axis, which is suggested to have arisen from the minor loss of Na and to be responsible for the emergence of the delicate superconductivity. Our findings resolve the controversies in NaAlSi and NaAlGe, while their similarities and differences provide an ideal opportunity to investigate a variety of exciting topological quantum states.
\end{abstract}

\maketitle

\section{\label{sec:level1}Introduction}
Nodal-line semimetal is a kind of topological quantum state with its valance and conducting bands crossing near the Fermi level in momentum space to form a closed loop \cite{fang2016topological}. These nodal lines are intrinsically related to a two-dimensional (2D) drumhead-like topological surface that resembles the edge states in nodal-point materials and holds the promise of realizing a series of magnetoelectric and electrical-transport properties. Because superconductors generally require a high electron concentration, nodal states are thought to be more favorable to harbor high $T_c$ superconductivity. The widely investigated prototype is PbTaSe$_2$ with its topological properties protected by the reflection symmetry and the breaking of space inversion symmetry \cite{ali2014noncentrosymmetric,bian2016topological,long2016single}. However, due to the existence of conventional bands near the Fermi level ($E_F$), whether the superconductivity is derived from its topological nature is still under debate \cite{guan2016superconducting,wang2016nodeless}.

NaAlSi has recently been discovered to be a new kind of nodal-line superconductor ($T_c$ = 7 K) \cite{kuroiwa2007superconductivity,yamada2021superconductivity} with clean topological bands around the Fermi level and does not intermingle with other bands, which means that the topological characteristics can be easy detected \cite{jin2019topological,yi2019topological,muechler2019superconducting,song2022spectroscopic}. NaAlSi crystallized in an anti-PbFCl layer structure with a space group of $No$.129 $P4/nmm$ \cite{kuroiwa2007superconductivity}, isostructural to LiFeAs \cite{tapp2008lifeas} and LiFeP \cite{deng2009new}. Al-Si is tetrahedrally coordinated and edge shared to form a quasi-2D layer, stacking along the c-axis with Na atoms in between. Theoretical calculations show multiple nodal lines in the vicinity of $E_F$ with the bands composed of a mixture of electrons from both Al and Si \cite{jin2019topological,yi2019topological,muechler2019superconducting}. The drumhead Fermi surface is contributed by the hole-like Si-3$p$ orbitals with nearly flat band dispersion and heavy electron effective mass, while the light electron bands mainly come from Al-3$s$ orbitals \cite{rhee2010naalsi}. The drumhead Fermi surfaces are suggested to be responsible for the intrinsic transport properties \cite{yamada2021superconductivity}, whereas the correspondence between superconductivity and topology is not clear. Intensive enthusiasms have been devoted to the research of NaAlSi recently. Muon spin relaxation ($\mu SR$) experiments confirmed the bulk superconductivity with a full gap opening \cite{muechler2019superconducting}. Its $T_c$ can be increased to 9 K under external pressure \cite{schoop2012effect}. Furthermore, a fractional superconducting signal has been observed on the surface of the single crystals aside from its bulk superconductivity \cite{hirai2022unusual}. 

It has long been known that NaAlSi has a sister compound NaAlGe by replacing Si for Ge \cite{kuroiwa2007superconductivity}. However, the powder form of NaAlGe does not show any trace of superconductivity down to the lowest temperature \cite{kuroiwa2007superconductivity} and its underlying mechanism is poorly understood \cite{rhee2010naalsi}. The band structures of both compounds are quite similar, except for a missing piece of a small electron pocket along the $\varGamma-M$ direction in NaAlGe \cite{rhee2010naalsi}. It has been suggested that the missing electron pocket is crucial for the absence of superconductivity in NaAlGe and is highly sensitive to the height of Si in NaAlSi. Other scenarios such as stronger electron-phonon coupling strength arising from the lighter Si or different electron pairing mechanism have been suggested \cite{rhee2010naalsi}. On the other hand, all the nodal lines are located slightly below $E_F$ \cite{wang2020rich}. Nontrivial phenomena can thus be anticipated if the system could be hole-doped to shift the Fermi level approaching the crossing sections.

\begin{figure*}[htbp] 
\centering 
\includegraphics[width=0.9\textwidth]{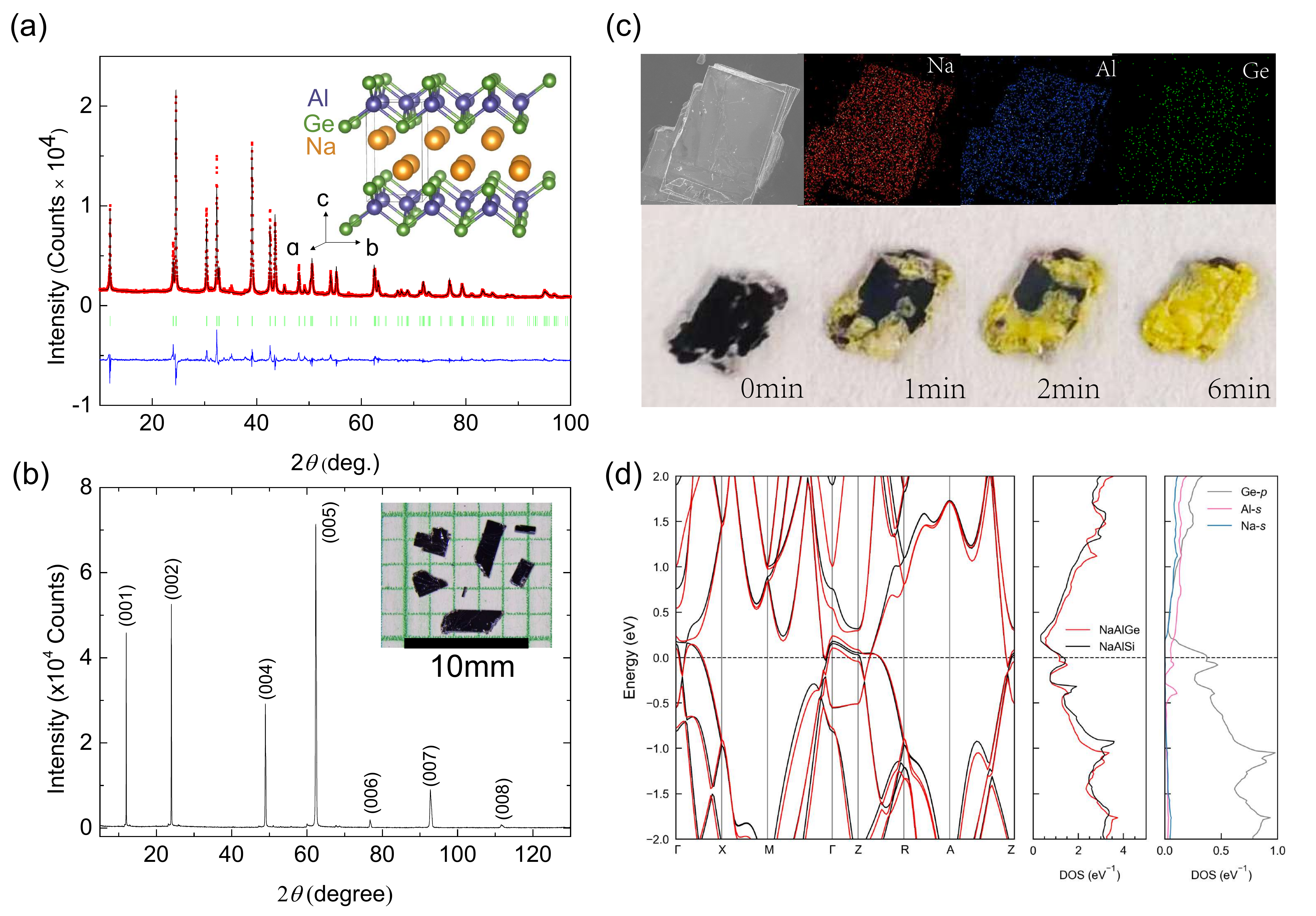}
\caption{Crystal structure, x-ray diffraction, composition, and electronic structure of NaAlGe. (a) Powder X-ray diffraction of NaAlGe. The red point represents the observed signal, and the black, green, and blue line represent the calculated curve, Bragg positions, and residual difference, respectively. Inset shows its crystal structure. (b) XRD pattern of NaAlGe $(00l)$ plane. Inset is an optical photo of the acquired single crystals. (c) The upper panel is an SEM image of NaAlGe and an EDS mapping of Na (red), Al (blue), and Ge (green). The lower panel displays the morphology of a crystal exposed to air over time. (d) The electronic structures of NaAlGe and NaAlSi.}
\label{FIG 1}
\end{figure*}

In this paper, we report the successful synthesis of NaAlGe single crystal, following the recent growth of NaAlSi single crystal by Na-Ga flux \cite{yamada2021superconductivity,hirai2022unusual}. The physical properties of NaAlGe single crystals are studied by electrical resistivity, magnetization, and heat capacity. Despite the intrinsic semimetallic/semiconducting ground state, NaAlGe single crystal shows a distinct superconducting transition with the Tc varies from 1.8 K to 3.5 K and a perfect diamagnetic response over 100 \%. Detailed characterizations reveal that the observed superconductivity is not a bulk response, but corresponds to a hole-doped Na$_{1-\delta}$AlGe with a slightly elongated $c$ axis. A conceptional illustration of a sandwich structure is given to explain all the observed phenomena. Our findings suggest that NaAlGe is an interesting platform to study the relationship between oxidation-modulated superconductivity and topology \cite{song2021competition,chen2021highly,yang2021discovery}.

\section{Experimental}
Powder crystals of NaAlGe were synthesized with Na lump, Al powder, and Ge powder as starting materials. The starting materials were weighted to a molar ratio of Na: Al: Ge = 1:1:1. The Al powder and Ge powder were well mixed and pressed into a pellet and loaded into alumina crucibles together with Na. The alumina crucibles covered with a cap were arc-sealed into steel cubes. We slowly heated these steel cubes to 850℃, held for 6h and furnace cooled to room temperature. A second annealing process was necessary to ensure homogeneity.

Single crystals were grown using Na lump, Al rod, Ge ingot, and Ga shot with a molar ratio of Na : Al : Ge : Ga = 3:1:1:0.5. We loaded the raw materials into an alumina crucible covered with a cap and sealed into a steel cube (SUS316). The sample was heated to 850℃, holding for 24h and then slowly cooling down to room temperature. To remove the Na-Ga flux, we put the crucible into anhydrous alcohol to remove the flux. Millimeter-size single crystals can be collected from the bottom of the crucible. Because of the highly hygroscopic nature of NaAlGe, all manipulations were performed in the Ar atmosphere. To minimize the degree of oxidation, the anhydrous alcohol was bubbled with argon to remove the trace amount of oxygen. We note that this washing process will still inevitably introduce surface oxidation. Thus, the intrinsic properties of NaAlGe are collected from the crystals directly pealed off from the flux in a glove box without alcohol treatment. For superconducting samples, hole-doping was realized by exposing the sample to air or soaking in alcohol for a few days.

Powder X-ray diffraction (XRD) patterns were performed on a Panalytical X’pert Diffractometer with a $Cu\,K_\alpha$ anode (1.5418{\text {Å}}) at room temperature. The Fullprof software was used for Rietveld refinement \cite{rodriguez2001fullprof}. The scanning electron microscopy (SEM) image of the single crystal was captured from the Hitachi S-4800 FE-SEM. the electron mapping of the sample was determined by the Energy Dispersive Spectroscopy (EDS). Transport, magnetic and thermodynamic properties were measured using a physical property measurement system (PPMS, Quantum Design), MPMS, and Keithley 2400-, SR830-based home-made electrical resistivity measurement system.

DFT calculations were performed in the Vienna \textit{ab initio} simulation package (VASP) \cite{kresse1996efficiency}. Generalized-gradient approximation (GGA) in the form of Pedrew-Burke-Ernzerhof (PBE) \cite{perdew1996generalized} was employed as the exchange-correlation functional. The projector-augmented-wave (PAW) pseudopotentials \cite{kresse1999ultrasoft} were used. In the calculation, the Brillouin zone is sampled in the $k$ space within the Monkhorst-Pack scheme \cite{monkhorst1976special}.

\begin{figure*}[htbp] 
\centering 
\includegraphics[width=0.8\textwidth]{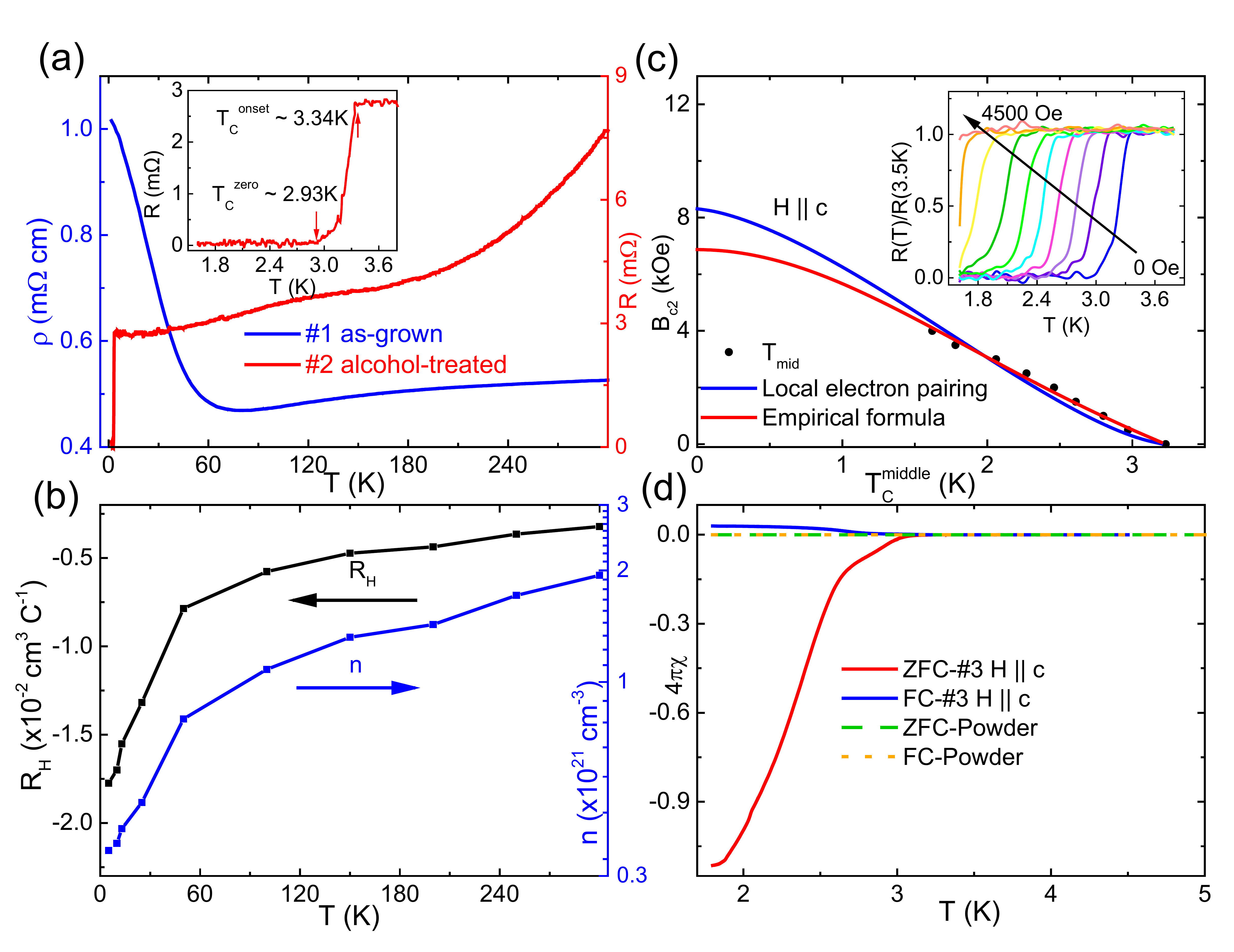} 
\caption{ Physical properties of the as-grown NaAlGe and alcohol-treated NaAlGe single crystals. (a) Electrical transport of as-grown NaAlGe (blue) and alcohol-treated NaAlGe (red) single crystals. Inset emphasizes the superconducting transition around 3.3 K. (b) Temperature dependences of Hall coefficient RH and carrier density n of the as-grown NaAlGe. (c) Temperature dependence of the upper critical field $H_{c2}$. The $T_c$ is defined at the midpoints of a 50\% drop of the normal-state resistance. The blue or red solid line represents the local electron pairing fitting or the empirical formula fitting. Inset shows the temperature-dependent resistance under different magnetic field along the $c$-axis. (d) Zero-field cooling (ZFC) and field cooling (FC) curves of NaAlGe powder and single crystal were measured under an external field of 10 Oe.}
\label{FIG 2} 
\end{figure*}

\section{Results and Discussion}
The NaAlGe powder was firstly synthesized in 2007 under high pressure \cite{kuroiwa2007superconductivity}. We found that NaAlGe powder can be feasibly synthesized at ambient pressure using steel tubes. Figure 1(a) shows the Rietveld refinement of the powder X-ray diffraction. The refined lattice constants were determined to be $a$ = $b$ = 4.1591(1){\text {Å}}, $c$ = 7.4164(2){\text {Å}}, while the reliable factors $R_p$ = 5.20\%, $R_{wp}$ = 7.21\%, $R_{exp}$ = 2.69\% and $\chi^2$ = 7.18. The inset of Fig. 1(a) shows the crystal structure of NaAlGe, isostructural to NaAlSi.  Figure 1(b) shows the $(00l)$ peak of the NaAlGe single crystal. As shown in the inset, all samples have shining surfaces. 

The upper panel of Fig. 1(c) shows the homogeneous dispersion of Na, Al, and Ge of the sample. Compared to the air-sensitive NaAlSi \cite{yamada2021superconductivity}, NaAlGe is much more reactive to moisture and highly hygroscopic. When exposed to air, there is an obvious color change on the crystal surface. As shown in the lower panel of Fig. 1(c), the crystal surface transforms from silvery white to light yellow and finally orange color. The final orange product may be a mixture of amorphous Na$_2$O$_2$ (which gives the color) and amorphous Al/Ge oxides. When reacted with water, both powder and single-crystal will burn or even explode, accompanied by an increase in the pH value of the remaining solvent. Actually, Si (1.90) has less electronegativity than that of Ge (2.01), although Si lies above Ge in the periodic table. It is thus not easy to understand why NaAlSi shows relatively higher stability. 

Figure 1(d) presents the electronic structures of NaAlSi and NaAlGe, which are consistent well with the previous reports \cite{jin2019topological,yi2019topological,muechler2019superconducting,wang2020rich}. The Al 3$s$ electron bands and the Ge/Si $4p/3p$ heavy hole bands dominate the Fermi surface, while the contribution of Na is small. The main dispersion crossing the Fermi surface happened in the plane, while the band along $\varGamma-Z$ is rather flat. This suggests the nature of two-dimensionality in NaAlGe. As a result, we find NaAlGe can be easily exfoliated. Detailed investigation of its transport behavior in reduced dimension is underway in our lab. It is worth noting that the electron pocket cannot be observed in NaAlSi along the $M-\varGamma$ path in our calculation, which is previously suggested to may be related to the superconductivity in NaAlSi \cite{rhee2010naalsi}. Many calculations \cite{muechler2019superconducting, jin2019topological, yi2019topological} and the recent magnetic torque measurement \cite{uji2022fermi} did not observe this electron pocket, as this pocket is sensitive to the height of Si. We suggest that the superconductivity observed NaAlGe originates from surface-oxidation-induced hole-doping, and will prove the electron pocket along $\varGamma-M$ is not of great importance (vide infra). The multiple nodal lines can be clearly distinguished slightly 20 $meV$ below the $E_F$.

To clarify the ground states of NaAlGe, we scrutinize the electrical transport and the magnetization behavior of NaAlGe single crystal. Due to the sensitivity to moisture, NaAlGe single crystal was carefully handled in the Ar-filled glove box to minimize the oxidation. The in-plane resistivity of the as-grown NaAlGe was shown in Figure 2(a). At a high temperature, the resistivity increases upon the temperature increases, indicating a metallic behavior. Below $ \sim $ 75 K, different from the metallic NaAlSi in the whole temperature range, the resistivity of NaAlGe exhibits an anomalous upturn. Our primary calculations including spin-orbit interaction failed to open a gap at the crossing point in both NaAlGe and NaAlSi, which excludes the possibility of a charge gap on the Fermi surface. NaAlGe possessed similar band structures but distinct electrical transport behaviors at low temperatures, indicating the ground state of NaAlGe may be unconventional. Hall coefficient and the corresponding carrier density are shown in Figure 2(b). There is a linear relationship between the Hall resistance and the magnetic field, which indicates the dominant carrier is electrons from the Al $3s$ band. At room temperature, the carrier density of NaAlGe is $2\times10^{21}$ $cm^{-3}$, but quickly drops below 50 K and reaches $3\times10^{20}$ $cm^{-3}$ at 5 K. The temperature dependence of carrier density in NaAlSi shows a dome at around 100 K \cite{yamada2021superconductivity}, while there is a monotonic increase with temperature in the carrier density of NaAlGe. In the whole temperature range, NaAlGe possessed a smaller carrier density than that of NaAlSi. The drop of carrier density at 50 K is consistent well with the resistivity increase shown in Fig. 2(a), pointing to the anomaly of resistivity that may originate from the loss of the carrier.

The physical properties of NaAlGe can be dramatically altered when alcohol was employed to remove the Na-Ga flux. The electrical transport is shown as the red solid line in Figure 2(a), which was totally different from that of the as-grown NaAlGe. The resistance behaves as a metal above 3.35 K. The resistance decreases rapidly above $ \sim $200 K and slowly below 200 K. When cooled to 3.3 K, it undergoes a superconducting transition. The $T_{c} ^{\mathrm{zero}}$ is 2.9 K and the transition width is around 0.4 K. To investigate the properties of the superconducting state in alcohol-treated NaAlGe, the magneto transport and magnetization were measured. The resistance at low temperatures under different magnetic fields along the $c$-axis was shown in the inset of Figure 2(c). $H_{c2}(T)$ was determined as the midpoint of the superconducting transition and summarized in Figure 2(c). To compare with NaAlSi, local electron pairing model \cite{RevModPhys.62.113}, $H_{c2} (T)=H_{c2} (0) {[1-(T/T_{c} )^{3/2} ]}^{3/2}$, was applied and the $H_{c2}$ (0) was fitted to be 6.8 kOe. Furthermore, an empirical formula ${\mu}_0 H_{c2}=H_{c2} (0)(1-t^2)/(1+t^2)$ gave an upper critical field of 8.3 kOe. Both were much less than Pauli limit 6 T and the $H_{c2}$(0) of NaAlSi \cite{yamada2021superconductivity}. The coherence length can be acquired from$ \mu_0 H_{c2} (0)=\Phi_0/2\pi \xi_0^2$ as 22 nm. Figure 2(d) plots the temperature dependence of the magnetic susceptibility of NaAlGe powder and alcohol-treated NaAlGe in the temperature range from 1.8 K to 5.0 K with a magnetic field along the $c$-axis. Our NaAlGe powder shows no superconductivity, as reported previously\cite{kuroiwa2007superconductivity}. Meanwhile, there are two obvious diamagnetic signals located at 3 K and 2.7 K in the zero-field cooling (ZFC) curve of the alcohol-treated NaAlGe single crystal, indicating the existence of multiple superconducting phases. At 1.8 K, the superconducting volume fraction is estimated to be 110$\%$. The part exceeding 100$\%$ may come from the incorrect demagnetization factor. 

\begin{figure}[htbp] 
\centering 
\includegraphics[width=0.5\textwidth]{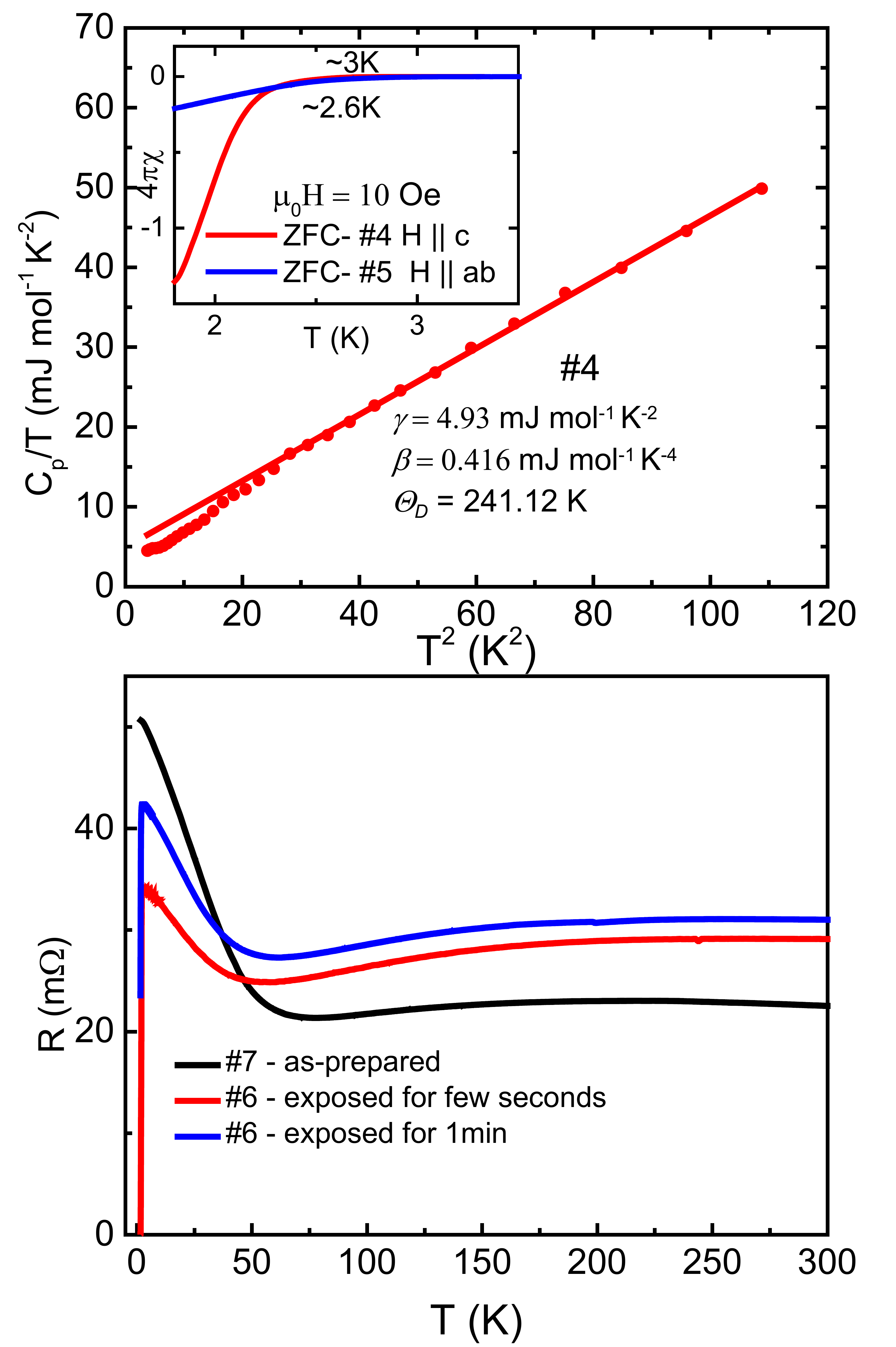} 
\caption{ (a) $C_p/T$ versus $T^2$ plot. The red solid line represents the conventional Debye model fitting. Inset shows the ZFC curve under a magnetic field of 10 Oe within the ab plane and along the $c$-axis. (c) Temperature dependence resistance with different exposure time in the air.} 
\label{FIG 3}
\end{figure}

To be rigorous, it is not enough to identify a new superconductor through electrical transport and magnetization. Heat capacity measurement, a method to analyze the entropy loss during the superconducting transition, is important to ensure the bulk feature \cite{carnicom2019importance}. We performed heat capacity measurement on an alcohol-treated NaAlGe with the superconducting transition at 2.6 K that determined by the ZFC curve (Fig. 3(a)). Although the superconducting volume fraction exceeded 100\% in magnetic susceptibility, no jump was observed on $C_p/T \sim  T^2$ curve at a corresponding transition temperature. By a fit of the conventional Debye model, the electron coefficient $\varGamma$ = 4.93 $mJ \cdot mol^{-1} \cdot K^2$ and the Debye temperature $\Theta_D$ = 241.12 K. Because alcohol treatment cannot influence the lattice and electron state, the $\varGamma$ and $\Theta_D$ should be similar to those of as-grown non-superconducting NaAlGe. Notably, there was a deviation from the linear relationship below 5 K in heat capacity, whose origin was not clear. The heat capacity indicates that the superconductivity should not be bulk form. In order to clarify the origin of superconductivity, the evolution of temperature-dependent resistance versus time was investigated. As mentioned above, the resistance undergoes a metal-semiconductor-like transition at about 50 K. We noticed that even for the as-grown crystal not being treated with alcohol, NaAlGe still undergoes a superconducting transition at low temperature after a slight explosion to air, as shown in Fig. 3(b). The superconductivity of the same crystal will gradually vanish over a long exposure time.

\begin{figure*}[htbp]
\centering 
\includegraphics[width=1.0\textwidth]{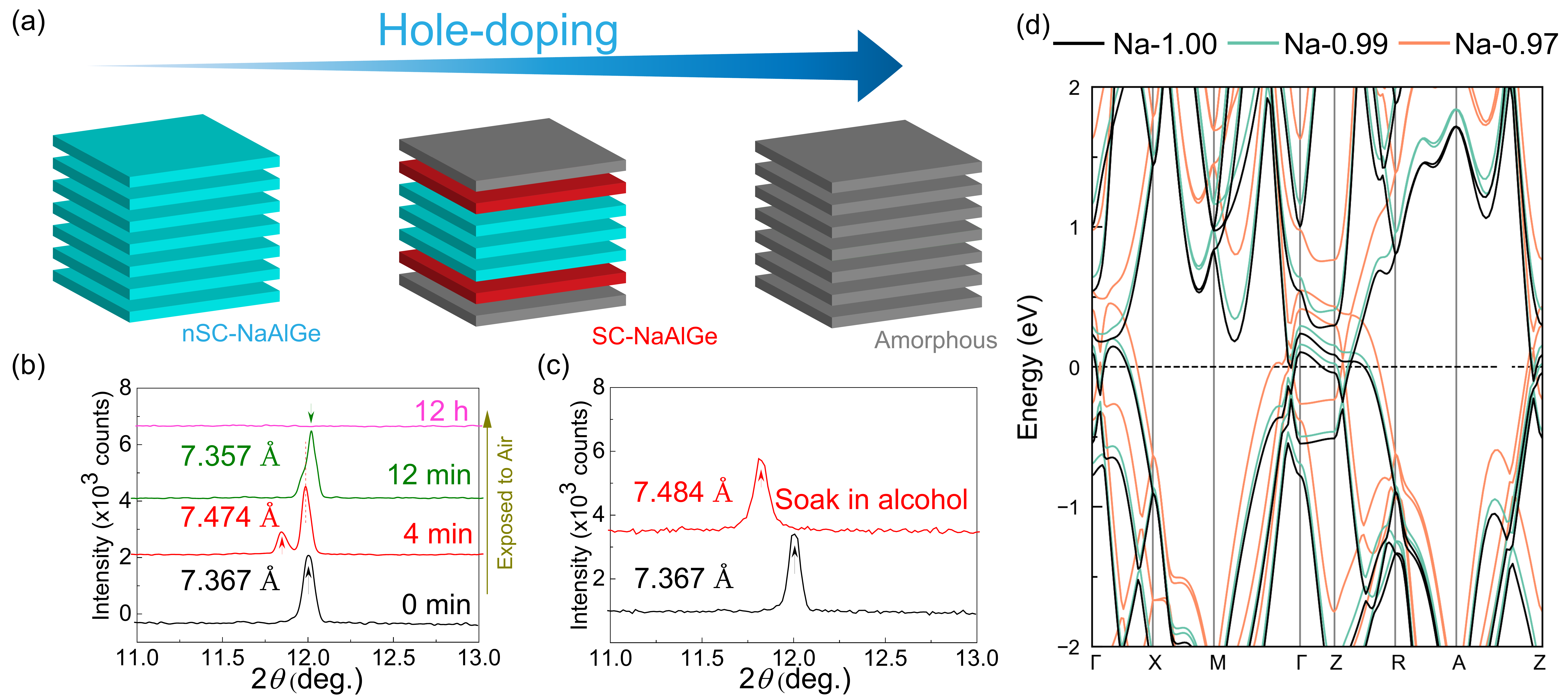}
\caption{ The origin of the superconductivity in NaAlGe single crystal. (a) The diagram of hole-doping induced superconductivity. The blue/brown layers represent the underdoped/over-doped layer. The red layer represents the hole-doping induced superconducting layer. (b) The (001) peak shift upon explosion in air. (c) The (001) peak shift before and after soaking in alcohol for several tens of days. (d) Electronic band structure of Na$_{1-\delta}$AlGe, $\delta$ = 0, 0.01, 0.03} 
\label{FIG 4} 
\end{figure*}

Various superconducting temperatures, the missing of a thermodynamic jump in heat capacity, and the evolution of superconductivity upon time all suggest that the superconductivity is not a bulk characteristic and the existence of a new superconducting phase. Inspired by the selective surface oxidation which introduces hole doping in CsV$_3$Sb$_5$ \cite{song2021competition} and the layered structure of NaAlGe, a hole-doping induced superconducting layer model was proposed. As shown in Fig. 4(a), the as-grown NaAlGe is non-superconducting. When exposed to air or trace water in the alcohol, the surface will be quickly over-doped. But the inside a few layers become hole-doped with a suitable doping content, i.e., oxidation when exposed to air or the loss of Na when soaked in alcohol. Due to the electromagnetic shielding effect, once two layers become superconducting, the remaining layers in between will exhibit an over-estimated superconducting diamagnetism and give a ‘fake’ perfect diamagnetization signal when the magnetic field is applied along the c-axis. On the contrary, when the magnetic field is within the $ab$-plane, it will show an actual superconducting volume fraction. This model explains the phenomenon that the superconducting volume fraction is always large under a magnetic field along the $c$-axis and small within the $ab$-plane, as shown in the inset Fig. 3(a). Another possibility is the development of a pseudogap  \cite{2022arXiv220604226Y} at around 50 K. Once the pseudogap is developed at a relatively high temperature, the entropy loss will be a gradual process and take place in a rather wide temperature range, leading to the absence of a jump in heat capacity. In such a case, the jump in heat capacity cannot be observed even if the majority of the layers reach optimal doping.

To further confirm this model, we trace the time evolution of the $(001)$ peak in x-ray for the sample exposed to air (Fig. 4(b)) and alcohol treatment (Fig. 4(c)). When exposed to air, continuous XRD measurements of the $(001)$ peak show an obvious peak shift from 12.00° to 11.83°, which corresponds to a c-axis change from 7.367 {\text {Å}} to 7.474 {\text {Å}}. As the increase of the explosion time, the shifted peak disappeared gradually and became a shoulder of the main (001) peak. The same situation happened in the alcohol-treated NaAlGe, whose c-axis changed from 7.367 {\text {Å}} to 7.484 {\text {Å}}. Because of the layered structure and the sensitivity of Na atoms between the Al-Ge layers, the peak shift suggests the loss of Na, which reduces the electric attractions and leads to the expansion of the $c$-axis. In this perspective, it is reasonable that alcohol treatment could achieve a larger $c$-axis, due to the higher efficiency of dissolved water in extracting Na atoms. Connected to the physical properties, Na loss corresponds to hole-doping. We attribute the superconductivity in alcohol-treated or air-exposed NaAlGe single crystals to hole-doped Na$_{1-\delta}$AlGe with a slightly larger $c$-axis. Although we cannot determine the a-axis of these transient new phases at present, we have good reason to believe it should have a smaller $a$-axis (\textless 4.1591 {\text {Å}}) than the as-prepared NaAlGe to maintain the nearly invariant volume of the unit cell. Moreover, we point out that the superconducting region in the hole-doped NaAlGe should be quite narrow as the powder NaAlGe does not superconduct, where the loss of Na is unavoidable. Such a loss of Na in the powder form may already push the system into an over-doped region. We noticed a slight shift peak of the $(001)$ peak to a higher angle after being exposed in air for 12 min in Fig 4(b), but it cannot be responsible for the superconductivity. It's because that the sample almost deteriorated, as shown in the lower panel of Fig 1(c).

The delicate superconductivity and the extremely narrow superconducting region in NaAlGe stem from the lack of Al-Ge binary compounds. When the loss of Na is large, the lattice will collapse and become amorphous phases without diffraction peaks. At the same time, the Na atoms become sodium oxides, possibly Na$_2$O$_2$, as evidenced in its characteristic yellow color shown in the low panel of Figure 1(c). We note that Al-Ge granular films have been reported to have a semiconductor/insulator-superconductor transition, with its highest $T_c$ lower than 1.8 K\cite{shapira1983semiconductor,gerber1997insulator,eytan1994resistivity}. What's more, there is no Na-Al binary compounds. NaGe and Na$_3$Ge are not superconductors. Thus, the superconductivity observed in NaAlGe is attributed to hole-doping rather than impurities.

It should be distinguished that the delicate superconductivity reported here is different from the general concept of filamentary superconductivity which has been reported in quasi-one-dimensional superconductors ZrTe$_3$ (low-temperature phase) \cite{nakajima1986anisotropic}, HfTe$_3$ \cite{li2017anisotropic} and several others, such as La$_{2-x}$Ba$_x$CuO$_4$\cite{moodenbaugh1988superconducting}, CaFe$_2$As$_2$ \cite{xiao2012evidence}, Ba(Fe,Co)$_2$As$_2$ \cite{xiao2012filamentary}, and the widely investigated K$_x$Fe$_{2-y}$Se$_2$ \cite{guo2010superconductivity,ding2013influence}. In these filamentary superconductors, their superconducting volume fraction cannot be further enhanced because of intrinsic phase separation or the observed superconductivity is merely an interface effect. However, in hole-doping NaAlGe, not only zero resistance state but also large diamagnetic drop can be observed, even if the magnetic field is along the ab-plane, in which case most electromagnetic shielding effect can be excluded for a layered structure. If the loss of Na can be adequately controlled in a gradual and mild manner, bulk superconductivity can be predicted to introduce a jump in heat capacity.

In order to trace the change of electronic structure, we performed the virtual crystal approximation (VCA) of Na$_{1-\delta}$AlGe, in which $\delta$ changes from 0 to 0.03. Because the amount of Na loss is small, the calculated results are reasonable and relable. Figure 4(d) shows the band structure of the Na$_{1-\delta}$AlGe, $\delta$ = 0, 0.01, 0.03. In the case of $\delta$ = 0, i.e. no loss of Na in NaAlGe, there is no electron pocket along $\varGamma - M$, as reported previously \cite{rhee2010naalsi,schoop2012effect,wang2020rich}. Upon hole-doping, the ``electron pocket" move far away from the Fermi surface. Thus, we exclude this ``electron pocket" to be responsible for the observed superconductivity in NaAlGe, as well as NaAlSi. As shown in Fig. 4(d), the crossing points of four nodal lines increase from below the Fermi surface to above that upon hole doping. The achieved superconductivity in a mildly hole-doped sample may correspond to a critical state when the crossing nodes sitting precisely on the $E_F$, which we call for further studies. Nevertheless, the topology is closely interwoven with superconductivity in the hole-doped NaAlGe, which makes it a promising candidate to realize the much-anticipated topological superconductivity.

\section{Conclusion}
In conclusion, we report the discovery of delicate superconductivity in slightly hole-doped NaAlGe single crystals with $T_c$ ranging from 2.3 K to 3.5 K. Hole-doping is realized by moderate oxidation or alcohol treatment. The over 100\% superconducting shielding fraction contradicts the lack of a thermodynamic jump in heat capacity, which inspires us to propose a sandwich structure to well explain all the observed phenomena. A new phase with a larger c-axis is responsible for the emergence of superconductivity, according to our time-dependent XRD pattern of (001) peak shift, while its precise crystal structure is unknown. The construction of a quantitative superconducting phase diagram in NaAlGe, as well as the explanation of the differences between it and its isostructural compound NaAlSi, merits further investigation. Our findings suggest that NaAlGe is an excellent platform to study the relationship between the modulated superconductivity and nodal lines on the Fermi surface.

~\\
\noindent \textbf{NOTE:} During the preparation of the manuscript, we noticed that T. Yamada \textit{et al.} also used Na-Ga flux to obtain the NaAlGe single crystals\cite{2022arXiv220604226Y}. While they are focusing on the intrinsic features of NaAlGe by evaporating Na flux to avoid oxidation, we intentionally expose the single crystal to the air or in alcohol to realize hole doping, and successfully introduce superconductivity in NaAlGe.

\begin{acknowledgments}
This work is financially supported by the MoST-Strategic International Cooperation in Science, Technology and Innovation Key Program (No. 2018YFE0202600) and National Key Research and Development Program of China (No. 2021YFA1401800).
\end{acknowledgments}

\nocite{*}
\bibliographystyle{apsrev4-2}
\bibliography{NaAlGe}

\end{document}